%% file: dym.tex
\documentstyle[12pt,epsfig]{article}
\newcommand{\be}{\begin{equation}}
\newcommand{\ee}{\end{equation}}
\newcommand{\ba}{\begin{eqnarray}}
\newcommand{\ea}{\end{eqnarray}}
\begin{document}
\renewcommand{\baselinestretch}{1.1}
\small\normalsize
\renewcommand{\theequation}{\arabic{section}.\arabic{equation}}
\renewcommand{\thesection}{\arabic{section}.}
\renewcommand{\thefootnote}{\fnsymbol{footnote}}
\language0

\begin{flushright}
{\sc BUHEP}-95-10\\March 1995
\end{flushright}

\thispagestyle{empty}

\vspace*{1.0cm}

\begin{center}

{\Large \bf Phase transition and dynamical-parameter method in U(1) 
gauge theory}

\vspace*{0.8cm}

{\bf Werner Kerler$^a$, Claudio Rebbi$^b$ and Andreas Weber$^a$}

\vspace*{0.3cm}

{\sl $^a$ Fachbereich Physik, Universit\"at Marburg, D-35032 Marburg, 
Germany\\ 
$^b$ Department of Physics, Boston University, Boston, MA 02215, USA}
\hspace*{3.6mm}

\end{center}

\vspace*{1.0cm}

\begin{abstract}
Monte Carlo simulations of the 4-dimensional compact U(1) lattice
gauge theory in the neighborhood of the transition point are made
difficult by the suppression of tunneling between the phases, which
becomes very strong as soon as the volume of the lattice grows to any
appreciable size.  This problem can be avoided by making the monopole
coupling a dynamical variable.  In this manner one can circumvent the
tunneling barrier by effectively riding on top of the peaks in the
energy distribution which meet for sufficiently large monopole
coupling.  Here we present an efficient method for determining the
parameters needed for this procedure, which can thus be implemented at
low computational cost also on large lattices.  This is particularly
important for a reliable determination of the transition point.  We
demonstrate the working of our method on a $16^4$ lattice.  We
obtain an equidistribution of configurations across the phase
transition even for such a relatively large lattice size.
\end{abstract}

\newpage

\section{Introduction}
\setcounter{equation}{0}
\hspace{3mm}
Recent investigations \cite{acg91,blsu92} of the compact U(1) lattice
gauge theory in 4 dimensions have produced energy histograms
indicative of a first-order transition.  In the corresponding Monte
Carlo simulations, however, the tunneling between the phases is
strongly suppressed. In order to overcome the difficulties due to
the lack of transitions the authors of \cite{acg91} introduce an
iterative reweighting for different $\beta$, while the authors of
\cite{blsu92} use a matching of hot and cold start results. The
problem is that on larger lattices conventional algorithms are not
able to induce transitions at all. Therefore, we have looked for
algorithmic alternatives.

 To reduce the slowing down of the Monte Carlo algorithm in systems
with a rough free-energy landscape the method of simulated tempering
\cite{mp92} has been proposed and applied to the random-field Ising
model. In this method the inverse temperature $\beta$ is promoted to
the status of dynamical variable, taking values which range over
some definite set.  In this manner one tries to utilize the fact that
at lower $\beta$ the free-energy barriers are lower. In an application
of this procedure to spin-glass simulations \cite{kr94} it has turned
out, however, that adjusting the set of temperatures and handling the
corresponding probability distribution in an efficient way is far from
straightforward. Nevertheless it has been possible to develop a
procedure \cite{kr94} leading to a reduction of slowing down
comparable to the one obtained with the multicanonical method
\cite{bc92}, and with the additional advantages of allowing full
vectorization of the code and of providing the canonical ensemble
directly.

In the Potts model in 2 dimensions, the strength of the first-order
transition decreases with the number of the degrees of freedom $q$ of
the spins, the transition becoming of second order for $q<5$.  This
has been used to set up an algorithm \cite{kw93} in which $q$ becomes
a dynamical variable: by opening the easier pathway along the
mountains of the joint probability distribution of $q$ and energy, one
avoids the need of relying, for large $q$, on the strongly suppressed
tunneling for equilibrating the configurations.  To implement
transitions between different $q$ cluster steps \cite{sw87} have been
inserted.  It turns out that by this algorithm one gains large factors
in the autocorrelation times also in comparison to the multicanonical
algorithm \cite{bn91}.

Proceeding along similar lines, we have obtained an efficient
algorithm for the U(1) gauge theory \cite{krw94}. We start from the
Wilson action supplemented by a monopole term \cite{bss85},
\be
S=\beta \sum_{\mu>\nu,x} (1-\cos \Theta_{\mu\nu,x})+
\lambda \sum_{\rho,x} |M_{\rho,x}| ,
\label{eq1}
\ee
where $M_{\rho,x}$ is the monopole content of 3 dimensional cubes
\cite{dt80}. One finds that the strength of the first order transition
decreases with $\lambda$, the transition ultimately becoming of second
order.  Thus, by making $\lambda$ a dynamical variable, we can again
dispose of the tunneling transitions and proceed instead along the
much easier pathway running over the top of the joint probability
distribution $P(E,\lambda)$, $E$ being the average plaquette
energy. With the use of appropriate Metropolis steps for the link
variables as well as $\lambda$, moreover, one can make the algorithm
fully vectorizable and parallelizable.

Before running the dynamical-parameter algorithm one has to determine
the position of the phase transition as function of $\lambda$ and some
parameters in the generalized action, which serve to enforce the
prescribed $\lambda$ distribution. On lattices of moderate size
(e.g.~$8^4$) this is relatively easy because there is still some
overlap between the peaks. On larger lattices determining these
quantities is much more difficult and it becomes then crucial to
perform the calculation without excessive computational cost. In the
present paper we develop a method to achieve this goal.  We
demonstrate its effectiveness illustrating results obtained for a
$16^4$ lattice. We will see that our method enables us to observe
transitions also on large lattices, which is very important for a
reliable determination of the transition point.

In Section 2 we outline the general features of the
dynamical-parameter method. In Section 3 we derive relations among
transition probabilities on which we will base the determination of
the quantities required for the implementation of the algorithm.  The
detailed procedure followed for their calculation is described in
Section 4.  In Section 5 we will present some numerical results.

\section{Outline of method}
\setcounter{equation}{0}
\hspace{3mm}
Conventional methods simulate the probability distribution
\be
\mu_{\lambda}(\Theta)= \exp(-S_{\lambda}(\Theta))/Z_{\lambda}
\ee
where $\lambda$ is a fixed parameter. In order to make $\lambda$ a
dynamical variable we consider $\mu_{\lambda}(\Theta)$ as the
conditioned probability to get a configuration $\Theta$ given a value
of $\lambda$ and prescribe a probability distribution $f(\lambda)$ to
get the joint probability distribution
$\mu(\Theta,\lambda)=f(\lambda)\mu_{\lambda}(\Theta)$.  To simulate
$\mu(\Theta,\lambda)$ we need it in the form
\be
\mu(\Theta,\lambda)=\exp(-S(\Theta,\lambda))/Z
\label{eq2}
\ee
where 
\be
S(\Theta,\lambda)=S_{\lambda}(\Theta)+g(\lambda) 
\label{eq3}
\ee
and 
\be
g(\lambda)=-\log(f(\lambda) Z/Z_{\lambda}).
\ee 
Eventually we will require that the values of $\lambda$ be approximately 
equidistributed, i.e. $f(\lambda) \approx$ const, which then gives
$g(\lambda) \approx \log Z_{\lambda}+$ const.

In our application of the algorithm each update of the link variables
$\Theta_{\mu,x}$ is followed by an update of $\lambda$, which can take
values from a discrete, ordered set $\lambda_q$ with $q=1,\ldots,n$.
The individual update steps are Metropolis steps in both cases. For
the $\lambda$ update we use a proposal matrix
$\frac{1}{2}(\delta_{q+1,q'}+\delta_{q,q'+1}+\delta_{q,1}
\delta_{q',1}+\delta_{q,n}\delta_{q',n})$ and an acceptance probability
$\min(1,\exp(S(\Theta,\lambda_{q})-S(\Theta,\lambda_{q'})))$.  The
above form of the proposal matrix implies that, if the current value
$\lambda_q$ is not extremal, then we choose as new candidate value for
$\lambda$ one of the two neighboring values, $\lambda_{q-1}$ or
$\lambda_{q+1}$, with equal probability, whereas, if $\lambda_q$ lies
at the boundary of the set of possible values, we preselect either its
(only) neighboring value or $\lambda_q$ itself, again with equal
probability.

In order to implement the simulation, one must fix $\beta(\lambda_q)$
(cfr.~(\ref{eq1})) and $g(\lambda_q)$ (cfr.~(\ref{eq2}) and
(\ref{eq3})), for all values of $q$. We demand
$\beta(\lambda_q)\approx \beta_w(\lambda_q)$, where $\beta_w$ is the
value of $\beta$ which makes both phases equally probable. Our
condition for fixing $g(\lambda_q)$ is $f(\lambda)\approx$ const. In
order to determine $\beta(\lambda_q)$ and $g(\lambda_q)$, we will use
the fact that in a simulation the transition probabilities between
neighboring values of $\lambda$ are very sensitive to these
quantities.

\section{Transition probabilities}
\setcounter{equation}{0}
\hspace{3mm}
 To derive relations which can be used for the envisaged determination of
$\beta(\lambda_q)$ and $g(\lambda_q)$ we use the probability for the
transition from a value $\lambda_q$ to a neighboring value $\lambda_{q'}$
\be
W(\Theta,q;q')=
\frac{1}{2}\min(1,\exp(S(\Theta,\lambda_{q})-S(\Theta,\lambda_{q'}))  
\label{tp}
\ee
and note that detailed balance implies
\be
f(\lambda_{q-1})\mu_{\lambda_{q-1}}(\Theta)W(\Theta,q-1;q)=
f(\lambda_q)\mu_{\lambda_q}(\Theta)W(\Theta,q;q-1) \; .
\label{db}
\ee

Let us consider subsets $K(q)$ of configurations $\Theta$
with probability distributions proportional to $\mu_{\lambda_q}(\Theta)$
and weight $w_K(q)=\sum_{\Theta \in K(q)} \mu_{\lambda_q}(\Theta)$.
If we introduce the average transition probability for the set $K(q)$ 
\be
p_K(q;q')=\frac{1}{w_K(q)}\sum_{\Theta \in K}\mu_{\lambda_q}(\Theta)
W(\Theta,q;q') \; ,
\label{atp}
\ee
by averaging (\ref{db}) we obtain
\be
f(\lambda_{q-1})\ w_K(q-1)\ p_{K}(q-1;q)=
f(\lambda_q)\ w_K(q)\ p_{K}(q;q-1) \; .
\label{adb}
\ee

We now apply (\ref{adb}) to sets $K_c$ and $K_h$ of configurations in the 
cold phase and in the hot phase separately. Because we are interested 
in cases where transitions between the phases are extremely rare, in 
practice it is easy to obtain sets of this type with numbers of
configurations sufficient for the present purpose. Also, for the
same reason, the corresponding equations can be considered to be 
independent. We assume that our two conditions, 
$\beta(\lambda)=\beta_w(\lambda)$ and 
$f(\lambda)=$const, are satisfied.  The condition on $\beta$
implies that the two phases are equally populated, i.e. 
$w_{K_c}=w_{K_h}$.  Moreover, since the two subsets $K_h$ and $K_c$
essentially exhaust the whole set of configurations (in the cases
we are considering the overlap is extremely small), all of the weights
are, to a very good approximation, equal to ${1 \over 2}$.  Using this
fact, the constancy of $f(\lambda)$, and (\ref{adb}) for $K=K_c$ 
and $K=K_h$ separately, we get a pair of equations which simplifies to
\ba
p_{K_c}(q-1;q)=p_{K_c}(q;q-1)\hspace{0.5mm}\nonumber\\
p_{K_h}(q-1;q)=p_{K_h}(q;q-1)
\label{trans}
\ea
This is what we will exploit to determine 
$\beta(\lambda_q)$ and $g(\lambda_q)$.

Our strategy will be to adjust $\beta(\lambda_q)$ and $g(\lambda_q)$, for
known $\beta(\lambda_{q-1})$ and $g(\lambda_{q-1})$, in such a way that
(\ref{trans}) holds.  Starting from given $\beta(\lambda_1)$ and
arbitrarily chosen $g(\lambda_1)$, in this manner we can obtain
$\beta(\lambda_q)$ and $g(\lambda_q)$ for $q=2,\ldots,n$.

\section{Determination of $\beta(\lambda)$ and $g(\lambda)$}
\setcounter{equation}{0}
\hspace{3mm}
 To begin our procedure we select a value for $\lambda_1$ in the region
where the peaks of the probability distribution associated to the two
phases strongly overlap so that tunneling occurs frequently and
$\beta(\lambda_1)$ can easily be obtained by a conventional
simulation. Because only the differences $g(\lambda_{q-1})-g(\lambda_q)$
are relevant we can choose $g(\lambda_1)$ arbitrarily.  Then for
$q=2,\ldots,n$ we consecutively determine $\beta(\lambda_q)$ and
$g(\lambda_q)$ for known $\beta(\lambda_{q-1})$ and $g(\lambda_{q-1})$.

In order to proceed from $q-1$ to $q$ we choose a new $\lambda_q$,
approximately at the same distance from $\lambda_{q-1}$ as in the previous
steps.  As a first rough approximation we obtain $\beta(\lambda_q)$ by
extrapolation from the former values.  At this point we use the sets
of cold and hot configurations $K_c(q-1)$ and $K_h(q-1)$ at
$\lambda_{q-1}$, available from the previous step, and generate two new
sets of $\Theta$ configurations $K_c(q)$ and $K_h(q)$ at $\lambda_q$ by
short Monte Carlo runs with cold and hot start, respectively.  
For each set $K_i(q')$ we can easily calculate the quantity 
\be
\tilde{p}_{K_i}(q';q'') =\frac{1}{N_{K_i(q')}} \sum_{\Theta \in K_i(q')}
W(\Theta,q';q'') \; , 
\ee 
where $N_{K_i(q')}$ is the number of configurations in the set and
$W(\Theta,q';q'')$ is given by (\ref{tp}) (of course, the variables
$q'$, $q''$ stand for $q-1$, $q$ or $q$, $q-1$, as appropriate).

Since $\tilde{p}$ approximates (\ref{atp}), this allows us to calculate
the quantities $p_{K_i}$ which, for the correct choice of 
$\beta(\lambda_q)$ and $g(\lambda_q)$, should satisfy (\ref{trans}). 
We adjust then $\beta(\lambda_q)$ and $g(\lambda_q)$ until the
equations (\ref{trans}) are satisfied.  In practice this takes
only a very small amount of computer time. We obtain good estimates 
for $\beta(\lambda_q)$ and $g(\lambda_q)$ though only approximate 
quantities enter (\ref{trans}) because the peaks related to
the phases vary only little with $\beta$. In addition, the quantities
$\tilde{p}_{K_i}(q';q'')$ are used to adjust the distances between
neighboring $\lambda$ values in such a way that one has roughly 
equal transition probabilities for all steps.

After a larger number of $q$ steps the errors may accumulate. Therefore
we perform short runs of the dynamical-parameter algorithm to test
whether the simulation does indeed travel along the mountains of the
distribution in the hot as well as in the cold phase. If it gets stuck
we slightly increase or decrease the couplings $\beta(\lambda_q)$ in the 
region of $\lambda$ where the transitions fail. We determine then the
corresponding values $g(\lambda_q)$ from the conditions
\be
\tilde{p}_{K_c}(q-1;q)+\tilde{p}_{K_h}(q-1;q)=
\tilde{p}_{K_c}(q;q-1)+\tilde{p}_{K_h}(q;q-1)
\label{tr}
\ee
and run the dynamical algorithm again. Typically one or two trials are 
sufficient.

After performing the simulations with dynamical $\lambda$, improved 
$\beta(\lambda_q)$ can be obtained by reweighting \cite{fs88} the 
distribution at the values of $\lambda$ where deviations from 
the equidistribution of configurations in the cold and hot phase
are seen to occur.  Corresponding new values for $g(\lambda_q)$ are 
then obtained from (\ref{tr}). Alternatively improved values for
$g(\lambda_q)$ can be obtained by replacing the current values with 
$g(\lambda_q)+\ln(f(\lambda_q))$.

\section{Numerical results}
\setcounter{equation}{0}
\hspace{3mm}
Our method has made it possible to determine the phase transition region
for a lattice as large as $16^4$ using only a moderate amount of
computer time.  We have used approximately $2\times 10^4$ sweeps per
each value of $\lambda$ to get the sets of configurations $K_c$ and
$K_h$ and approximately $4\times 10^4$ sweeps, in total, in the short
test runs with the dynamical algorithm.  These preliminary calculations
have been used to determine $\lambda_q$, $\beta(\lambda_q)$ and
$g(\lambda_q)$ following the procedure described in Section 4.  Our
results for these parameters are reproduced in Table 1.  Altogether we
used 25 values for $\lambda$ ranging from $\lambda=0.4$ (our starting
point) down to $\lambda=0$.  In our simulations with dynamical $\lambda$
we performed approximately $10^6$ sweeps of the lattice and we observed
a large number of transitions between the phases also on the $16^4$
lattice.

We define as location $\beta_{\mbox{\scriptsize{C}}}$ of the phase
transition the maximum of the specific heat. We have used reweighting
\cite{fs88} in order to explore a range of $\beta$ in the neighborhood
of the value $\beta(\lambda_q)$.  As a matter of fact reweighting is
necessary not only to find $\beta_{\mbox{\scriptsize{C}}}$, but also to
determine accurately $\beta_w$  (the value of $\beta$ where the
configurations are equidistributed between the phases) since in order to
implement the procedure of Section 4 we only needed to make the areas
under the peaks approximately equal.

Figure 1 shows $\beta_{\mbox{\scriptsize{C}}}$ as function of $\lambda$
for the $16^4$ lattice and also our earlier results \cite{krw94} for the
$8^4$ lattice. In particular, for the $16^4$ lattice at $\lambda=0$, we
obtain the value $\beta_{\mbox{\scriptsize{C}}}=1.01084(5)$ , where the
error has been estimated from the fluctuation of different samples. This
confirms the result $\beta_{\mbox{\scriptsize{C}}}=1.01082(6)$ obtained
in Ref.~\cite{blsu92} by a matching procedure.

In Figure 2 we show the distribution $P(E,\lambda)$ at $\beta_w$ (for
$\lambda=0.6$ at $\beta_C$) which we got (after reweighting) from our
simulations. The data have been obtained with the dynamical-parameter
algorithm, except for $\lambda=0.5$ and $\lambda=0.6$ where the peaks
overlap and the conventional Metropolis algorithm is adequate. Comparing
with the corresponding figure for the $8^4$ lattice in \cite{krw94} the
much stronger suppression of tunneling in the transition region is
obvious.  In fact, on the $8^4$ lattice, because of the overlap between
the peaks, there is still substantial tunneling.  (For this reason, in
our earlier simulations on the $8^4$ lattice we could determine
$g(\lambda_q)$ following the less sophisticated procedure based on
(\ref{tr}).)

In regards to the efficiency of our algorithm versus conventional
methods, the number of sweeps required to observe comparable numbers of
tunnelings is greatly reduced already on an $8^4$ lattice.  One must
make here a distinction (cfr. the discussion in \cite{krw94}) according to
whether one is interested in all values of $\lambda$, in which case
our method produces all of the results in one stroke, or in a single
$\lambda$.  In the latter case, since our method requires that one still
simulates a whole range of $\lambda$ values, fairness requires that the
the observed mean time between tunnelings be multiplied roughly by the
number of $\lambda$ values considered.  Even in this case, with an $8^4$
lattice there is still considerable gain, for example, for
$\lambda=-0.3$, and some gain remains also for $\lambda=0$ \cite{krw94}.

With a $16^4$ lattice a comparison is, as a matter of fact, impossible,
simply because the separation between the phases is so strong that with
conventional algorithms one does not observe any transition at all.
With our algorithm, instead, on a $16^4$ lattice and for $\lambda=0$, we
observe average tunneling times of the order of $10^3$ (for tunneling
time we follow the definition of \cite{bp91}). If we were interested
only in $\lambda=0$, this number ought to be multiplied by 25, i.e. the
number of $\lambda_q$ involved. This is certainly not a small time,
however it is small as compared to infinity, which corresponds to
observing no transition at all.

For a further reduction of the autocorrelation times, in addition to
circumventing tunneling, one would have to replace the local Metropolis
steps for $\Theta$ with more efficient ones. In the dynamical parameter
algorithm for the Potts model \cite{kw93} the cluster steps, which were
originally introduced to implement the transitions between different
$q$, have the additional advantage of reducing critical slowing down
and, correspondingly, the autocorrelation time in the second order
region.  However, at this stage an implementation of cluster steps for
gauge theories with continous groups appears very problematic, if not
plainly impossible \cite{km94}.  A more promising direction to pursue
might be along the lines of multi-scale algorithms \cite{ab91}, provided
that these could be modified to account for the actual structure of the
configurations.

\section*{Acknowledgements}
\hspace{3mm}
One of us (W.K.) wishes to thank the Physics Department of Boston University 
for kind hospitality during his visits. 
This research was supported in part under DFG grants Ke 250/7-2 and 250/12-1 
and under DOE grant DE-FG02-91ER40676.
The computations were done on the CM5 of the Center for Computational
Science of Boston University and on the CM5 of the GMD at St.~Augustin.

\newpage

\begin{center}

{\bf Table 1}
\vspace{5mm}

$\beta(\lambda)$ and $g(\lambda)$ of simulations on $16^4$ lattice
\vspace{5mm}

\begin{tabular}{|c|c|c|}
\hline
$\lambda$ &  $\beta$ & $g(\lambda)$ \\
\hline
  0.000 &   1.01078  &  $0.0000000\times 10^{0}$ \\
  0.020 &   0.99941  &  $1.3184682\times 10^{3}$ \\
  0.040 &   0.98800  &  $2.6584643\times 10^{3}$ \\
  0.060 &   0.97656  &  $4.0189975\times 10^{3}$ \\
  0.080 &   0.96509  &  $5.4003278\times 10^{3}$ \\
  0.100 &   0.95358  &  $6.8044118\times 10^{3}$ \\
  0.120 &   0.94206  &  $8.2270033\times 10^{3}$ \\
  0.140 &   0.93049  &  $9.6746392\times 10^{3}$ \\
  0.160 &   0.91888  &  $1.1146171\times 10^{4}$ \\
  0.180 &   0.90722  &  $1.2643661\times 10^{4}$ \\
  0.200 &   0.89551  &  $1.4167621\times 10^{4}$ \\
  0.215 &   0.88670  &  $1.5327348\times 10^{4}$ \\
  0.230 &   0.87785  &  $1.6504340\times 10^{4}$ \\
  0.245 &   0.86897  &  $1.7697237\times 10^{4}$ \\
  0.260 &   0.86006  &  $1.8906291\times 10^{4}$ \\
  0.275 &   0.85111  &  $2.0133296\times 10^{4}$ \\
  0.290 &   0.84211  &  $2.1380201\times 10^{4}$ \\
  0.300 &   0.83610  &  $2.2219762\times 10^{4}$ \\
  0.315 &   0.82701  &  $2.3501546\times 10^{4}$ \\
  0.330 &   0.81790  &  $2.4799088\times 10^{4}$ \\
  0.345 &   0.80870  &  $2.6124186\times 10^{4}$ \\
  0.360 &   0.79945  &  $2.7470739\times 10^{4}$ \\
  0.375 &   0.79015  &  $2.8839072\times 10^{4}$ \\
  0.390 &   0.78080  &  $3.0229562\times 10^{4}$ \\
  0.400 &   0.77455  &  $3.1167136\times 10^{4}$ \\
\hline
\end{tabular}

\end{center}

\newpage

\begin{figure}
\input fig1.tex
\caption{Location of phase transition as function of
$\lambda$ for $8^4$ (circles) and $16^4$ (crosses) lattices.}
\end{figure}

\begin{figure}[tb]
\begin{center}
\leavevmode
\psfig{figure=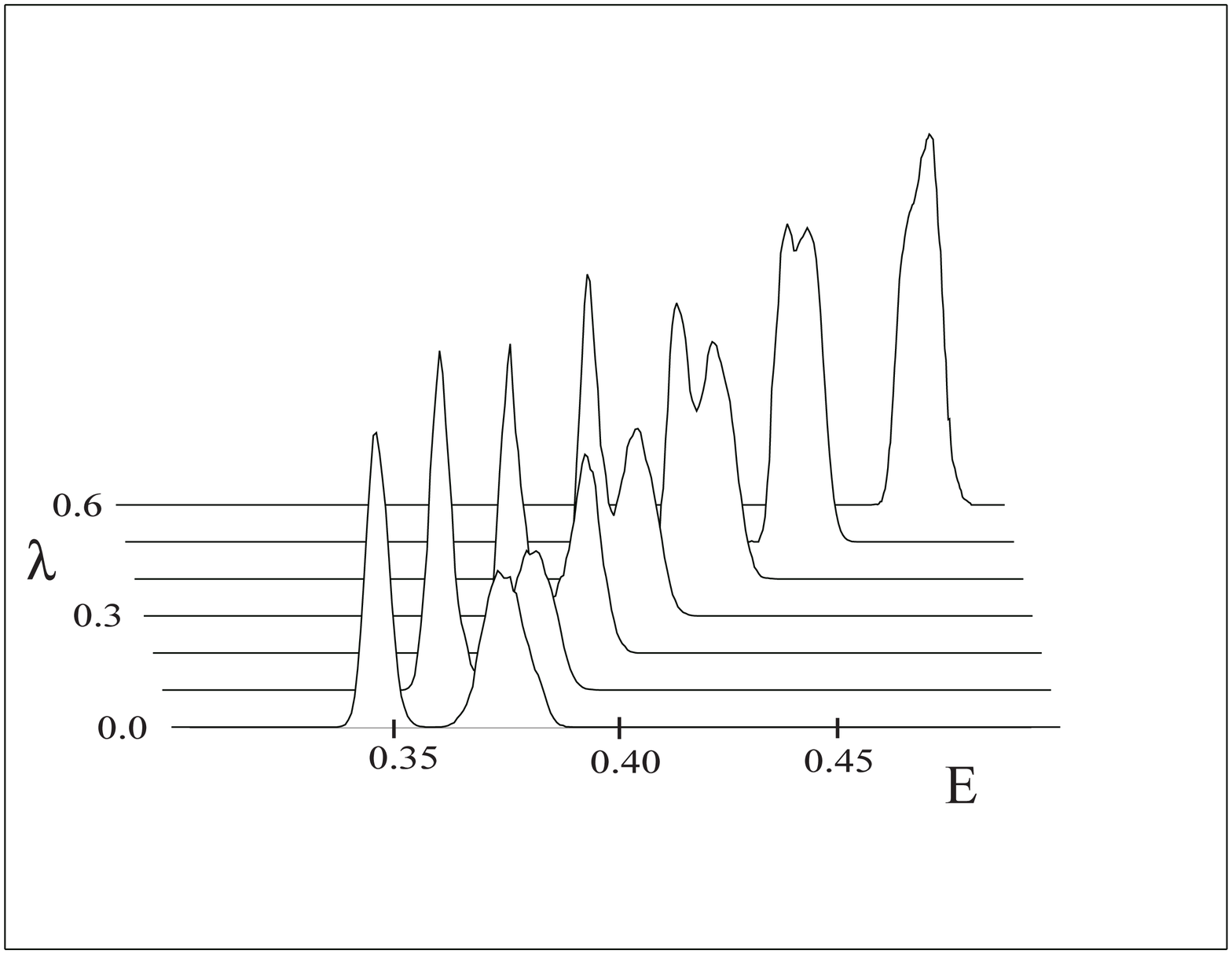,bbllx=2011bp,bblly=2080bp,bburx=2731bp,%
bbury=2637bp,width=10cm,angle=0}
\end{center}
\vspace{1cm}
\caption{Distribution $P(E,\lambda)$ at
the phase transition line on $16^4$ lattice.}
\end{figure}

\end{document}

%% file: fig1.tex
% GNUPLOT: LaTeX picture
\setlength{\unitlength}{0.240900pt}
\ifx\plotpoint\undefined\newsavebox{\plotpoint}\fi
\sbox{\plotpoint}{\rule[-0.200pt]{0.400pt}{0.400pt}}%
\begin{picture}(1500,900)(0,0)
\font\gnuplot=cmr10 at 10pt
\gnuplot
\sbox{\plotpoint}{\rule[-0.200pt]{0.400pt}{0.400pt}}%
\put(220.0,113.0){\rule[-0.200pt]{4.818pt}{0.400pt}}
\put(198,113){\makebox(0,0)[r]{0.6}}
\put(1416.0,113.0){\rule[-0.200pt]{4.818pt}{0.400pt}}
\put(220.0,334.0){\rule[-0.200pt]{4.818pt}{0.400pt}}
\put(198,334){\makebox(0,0)[r]{0.8}}
\put(1416.0,334.0){\rule[-0.200pt]{4.818pt}{0.400pt}}
\put(220.0,555.0){\rule[-0.200pt]{4.818pt}{0.400pt}}
\put(198,555){\makebox(0,0)[r]{1.0}}
\put(1416.0,555.0){\rule[-0.200pt]{4.818pt}{0.400pt}}
\put(220.0,777.0){\rule[-0.200pt]{4.818pt}{0.400pt}}
\put(198,777){\makebox(0,0)[r]{1.2}}
\put(1416.0,777.0){\rule[-0.200pt]{4.818pt}{0.400pt}}
\put(331.0,113.0){\rule[-0.200pt]{0.400pt}{4.818pt}}
\put(331,68){\makebox(0,0){-0.3}}
\put(331.0,812.0){\rule[-0.200pt]{0.400pt}{4.818pt}}
\put(662.0,113.0){\rule[-0.200pt]{0.400pt}{4.818pt}}
\put(662,68){\makebox(0,0){0.0}}
\put(662.0,812.0){\rule[-0.200pt]{0.400pt}{4.818pt}}
\put(994.0,113.0){\rule[-0.200pt]{0.400pt}{4.818pt}}
\put(994,68){\makebox(0,0){0.3}}
\put(994.0,812.0){\rule[-0.200pt]{0.400pt}{4.818pt}}
\put(1325.0,113.0){\rule[-0.200pt]{0.400pt}{4.818pt}}
\put(1325,68){\makebox(0,0){0.6}}
\put(1325.0,812.0){\rule[-0.200pt]{0.400pt}{4.818pt}}
\put(220.0,113.0){\rule[-0.200pt]{292.934pt}{0.400pt}}
\put(1436.0,113.0){\rule[-0.200pt]{0.400pt}{173.207pt}}
\put(220.0,832.0){\rule[-0.200pt]{292.934pt}{0.400pt}}
\put(45,472){\makebox(0,0){$\beta_c$}}
\put(828,23){\makebox(0,0){$\lambda$}}
\put(828,877){\makebox(0,0){ }}
\put(1436,666){\makebox(0,0)[l]{ }}
\put(220.0,113.0){\rule[-0.200pt]{0.400pt}{173.207pt}}
\put(331,753){\circle{12}}
\put(386,722){\circle{12}}
\put(441,690){\circle{12}}
\put(496,659){\circle{12}}
\put(552,627){\circle{12}}
\put(607,595){\circle{12}}
\put(662,564){\circle{12}}
\put(717,532){\circle{12}}
\put(773,500){\circle{12}}
\put(828,467){\circle{12}}
\put(883,435){\circle{12}}
\put(939,402){\circle{12}}
\put(994,368){\circle{12}}
\put(1049,334){\circle{12}}
\put(1104,299){\circle{12}}
\put(1160,263){\circle{12}}
\put(1215,227){\circle{12}}
\put(1243,208){\circle{12}}
\put(1270,189){\circle{12}}
\put(1298,170){\circle{12}}
\put(1325,150){\circle{12}}
\put(331,753){\usebox{\plotpoint}}
\put(321.0,753.0){\rule[-0.200pt]{4.818pt}{0.400pt}}
\put(321.0,753.0){\rule[-0.200pt]{4.818pt}{0.400pt}}
\put(386.0,721.0){\usebox{\plotpoint}}
\put(376.0,721.0){\rule[-0.200pt]{4.818pt}{0.400pt}}
\put(376.0,722.0){\rule[-0.200pt]{4.818pt}{0.400pt}}
\put(441,690){\usebox{\plotpoint}}
\put(431.0,690.0){\rule[-0.200pt]{4.818pt}{0.400pt}}
\put(431.0,690.0){\rule[-0.200pt]{4.818pt}{0.400pt}}
\put(496.0,658.0){\usebox{\plotpoint}}
\put(486.0,658.0){\rule[-0.200pt]{4.818pt}{0.400pt}}
\put(486.0,659.0){\rule[-0.200pt]{4.818pt}{0.400pt}}
\put(552,627){\usebox{\plotpoint}}
\put(542.0,627.0){\rule[-0.200pt]{4.818pt}{0.400pt}}
\put(542.0,627.0){\rule[-0.200pt]{4.818pt}{0.400pt}}
\put(607.0,595.0){\usebox{\plotpoint}}
\put(597.0,595.0){\rule[-0.200pt]{4.818pt}{0.400pt}}
\put(597.0,596.0){\rule[-0.200pt]{4.818pt}{0.400pt}}
\put(662,564){\usebox{\plotpoint}}
\put(652.0,564.0){\rule[-0.200pt]{4.818pt}{0.400pt}}
\put(652.0,564.0){\rule[-0.200pt]{4.818pt}{0.400pt}}
\put(717,532){\usebox{\plotpoint}}
\put(707.0,532.0){\rule[-0.200pt]{4.818pt}{0.400pt}}
\put(707.0,532.0){\rule[-0.200pt]{4.818pt}{0.400pt}}
\put(773,500){\usebox{\plotpoint}}
\put(763.0,500.0){\rule[-0.200pt]{4.818pt}{0.400pt}}
\put(763.0,500.0){\rule[-0.200pt]{4.818pt}{0.400pt}}
\put(828,467){\usebox{\plotpoint}}
\put(818.0,467.0){\rule[-0.200pt]{4.818pt}{0.400pt}}
\put(818.0,467.0){\rule[-0.200pt]{4.818pt}{0.400pt}}
\put(883,435){\usebox{\plotpoint}}
\put(873.0,435.0){\rule[-0.200pt]{4.818pt}{0.400pt}}
\put(873.0,435.0){\rule[-0.200pt]{4.818pt}{0.400pt}}
\put(939.0,401.0){\usebox{\plotpoint}}
\put(929.0,401.0){\rule[-0.200pt]{4.818pt}{0.400pt}}
\put(929.0,402.0){\rule[-0.200pt]{4.818pt}{0.400pt}}
\put(994,368){\usebox{\plotpoint}}
\put(984.0,368.0){\rule[-0.200pt]{4.818pt}{0.400pt}}
\put(984.0,368.0){\rule[-0.200pt]{4.818pt}{0.400pt}}
\put(1049,334){\usebox{\plotpoint}}
\put(1039.0,334.0){\rule[-0.200pt]{4.818pt}{0.400pt}}
\put(1039.0,334.0){\rule[-0.200pt]{4.818pt}{0.400pt}}
\put(1104,299){\usebox{\plotpoint}}
\put(1094.0,299.0){\rule[-0.200pt]{4.818pt}{0.400pt}}
\put(1094.0,299.0){\rule[-0.200pt]{4.818pt}{0.400pt}}
\put(1160,263){\usebox{\plotpoint}}
\put(1150.0,263.0){\rule[-0.200pt]{4.818pt}{0.400pt}}
\put(1150.0,263.0){\rule[-0.200pt]{4.818pt}{0.400pt}}
\put(1215,227){\usebox{\plotpoint}}
\put(1205.0,227.0){\rule[-0.200pt]{4.818pt}{0.400pt}}
\put(1205.0,227.0){\rule[-0.200pt]{4.818pt}{0.400pt}}
\put(1243,208){\usebox{\plotpoint}}
\put(1233.0,208.0){\rule[-0.200pt]{4.818pt}{0.400pt}}
\put(1233.0,208.0){\rule[-0.200pt]{4.818pt}{0.400pt}}
\put(1270,189){\usebox{\plotpoint}}
\put(1260.0,189.0){\rule[-0.200pt]{4.818pt}{0.400pt}}
\put(1260.0,189.0){\rule[-0.200pt]{4.818pt}{0.400pt}}
\put(1298,170){\usebox{\plotpoint}}
\put(1288.0,170.0){\rule[-0.200pt]{4.818pt}{0.400pt}}
\put(1288.0,170.0){\rule[-0.200pt]{4.818pt}{0.400pt}}
\put(1325,150){\usebox{\plotpoint}}
\put(1315.0,150.0){\rule[-0.200pt]{4.818pt}{0.400pt}}
\put(1315.0,150.0){\rule[-0.200pt]{4.818pt}{0.400pt}}
\put(662,567){\makebox(0,0){$\times$}}
\put(684,555){\makebox(0,0){$\times$}}
\put(706,542){\makebox(0,0){$\times$}}
\put(729,530){\makebox(0,0){$\times$}}
\put(751,517){\makebox(0,0){$\times$}}
\put(773,504){\makebox(0,0){$\times$}}
\put(795,491){\makebox(0,0){$\times$}}
\put(817,479){\makebox(0,0){$\times$}}
\put(839,466){\makebox(0,0){$\times$}}
\put(861,453){\makebox(0,0){$\times$}}
\put(883,440){\makebox(0,0){$\times$}}
\put(900,430){\makebox(0,0){$\times$}}
\put(916,420){\makebox(0,0){$\times$}}
\put(933,411){\makebox(0,0){$\times$}}
\put(950,401){\makebox(0,0){$\times$}}
\put(966,391){\makebox(0,0){$\times$}}
\put(983,381){\makebox(0,0){$\times$}}
\put(994,374){\makebox(0,0){$\times$}}
\put(1010,364){\makebox(0,0){$\times$}}
\put(1027,354){\makebox(0,0){$\times$}}
\put(1044,344){\makebox(0,0){$\times$}}
\put(1060,334){\makebox(0,0){$\times$}}
\put(1077,323){\makebox(0,0){$\times$}}
\put(1093,313){\makebox(0,0){$\times$}}
\put(1104,306){\makebox(0,0){$\times$}}
\put(1215,235){\makebox(0,0){$\times$}}
\put(1325,160){\makebox(0,0){$\times$}}
\put(662,567){\usebox{\plotpoint}}
\put(652.0,567.0){\rule[-0.200pt]{4.818pt}{0.400pt}}
\put(652.0,567.0){\rule[-0.200pt]{4.818pt}{0.400pt}}
\put(684,555){\usebox{\plotpoint}}
\put(674.0,555.0){\rule[-0.200pt]{4.818pt}{0.400pt}}
\put(674.0,555.0){\rule[-0.200pt]{4.818pt}{0.400pt}}
\put(706,542){\usebox{\plotpoint}}
\put(696.0,542.0){\rule[-0.200pt]{4.818pt}{0.400pt}}
\put(696.0,542.0){\rule[-0.200pt]{4.818pt}{0.400pt}}
\put(729.0,529.0){\usebox{\plotpoint}}
\put(719.0,529.0){\rule[-0.200pt]{4.818pt}{0.400pt}}
\put(719.0,530.0){\rule[-0.200pt]{4.818pt}{0.400pt}}
\put(751,517){\usebox{\plotpoint}}
\put(741.0,517.0){\rule[-0.200pt]{4.818pt}{0.400pt}}
\put(741.0,517.0){\rule[-0.200pt]{4.818pt}{0.400pt}}
\put(773,504){\usebox{\plotpoint}}
\put(763.0,504.0){\rule[-0.200pt]{4.818pt}{0.400pt}}
\put(763.0,504.0){\rule[-0.200pt]{4.818pt}{0.400pt}}
\put(795,491){\usebox{\plotpoint}}
\put(785.0,491.0){\rule[-0.200pt]{4.818pt}{0.400pt}}
\put(785.0,491.0){\rule[-0.200pt]{4.818pt}{0.400pt}}
\put(817,479){\usebox{\plotpoint}}
\put(807.0,479.0){\rule[-0.200pt]{4.818pt}{0.400pt}}
\put(807.0,479.0){\rule[-0.200pt]{4.818pt}{0.400pt}}
\put(839,466){\usebox{\plotpoint}}
\put(829.0,466.0){\rule[-0.200pt]{4.818pt}{0.400pt}}
\put(829.0,466.0){\rule[-0.200pt]{4.818pt}{0.400pt}}
\put(861,453){\usebox{\plotpoint}}
\put(851.0,453.0){\rule[-0.200pt]{4.818pt}{0.400pt}}
\put(851.0,453.0){\rule[-0.200pt]{4.818pt}{0.400pt}}
\put(883,440){\usebox{\plotpoint}}
\put(873.0,440.0){\rule[-0.200pt]{4.818pt}{0.400pt}}
\put(873.0,440.0){\rule[-0.200pt]{4.818pt}{0.400pt}}
\put(900,430){\usebox{\plotpoint}}
\put(890.0,430.0){\rule[-0.200pt]{4.818pt}{0.400pt}}
\put(890.0,430.0){\rule[-0.200pt]{4.818pt}{0.400pt}}
\put(916,420){\usebox{\plotpoint}}
\put(906.0,420.0){\rule[-0.200pt]{4.818pt}{0.400pt}}
\put(906.0,420.0){\rule[-0.200pt]{4.818pt}{0.400pt}}
\put(933.0,410.0){\usebox{\plotpoint}}
\put(923.0,410.0){\rule[-0.200pt]{4.818pt}{0.400pt}}
\put(923.0,411.0){\rule[-0.200pt]{4.818pt}{0.400pt}}
\put(950,401){\usebox{\plotpoint}}
\put(940.0,401.0){\rule[-0.200pt]{4.818pt}{0.400pt}}
\put(940.0,401.0){\rule[-0.200pt]{4.818pt}{0.400pt}}
\put(966,391){\usebox{\plotpoint}}
\put(956.0,391.0){\rule[-0.200pt]{4.818pt}{0.400pt}}
\put(956.0,391.0){\rule[-0.200pt]{4.818pt}{0.400pt}}
\put(983,381){\usebox{\plotpoint}}
\put(973.0,381.0){\rule[-0.200pt]{4.818pt}{0.400pt}}
\put(973.0,381.0){\rule[-0.200pt]{4.818pt}{0.400pt}}
\put(994,374){\usebox{\plotpoint}}
\put(984.0,374.0){\rule[-0.200pt]{4.818pt}{0.400pt}}
\put(984.0,374.0){\rule[-0.200pt]{4.818pt}{0.400pt}}
\put(1010,364){\usebox{\plotpoint}}
\put(1000.0,364.0){\rule[-0.200pt]{4.818pt}{0.400pt}}
\put(1000.0,364.0){\rule[-0.200pt]{4.818pt}{0.400pt}}
\put(1027,354){\usebox{\plotpoint}}
\put(1017.0,354.0){\rule[-0.200pt]{4.818pt}{0.400pt}}
\put(1017.0,354.0){\rule[-0.200pt]{4.818pt}{0.400pt}}
\put(1044,344){\usebox{\plotpoint}}
\put(1034.0,344.0){\rule[-0.200pt]{4.818pt}{0.400pt}}
\put(1034.0,344.0){\rule[-0.200pt]{4.818pt}{0.400pt}}
\put(1060,334){\usebox{\plotpoint}}
\put(1050.0,334.0){\rule[-0.200pt]{4.818pt}{0.400pt}}
\put(1050.0,334.0){\rule[-0.200pt]{4.818pt}{0.400pt}}
\put(1077,323){\usebox{\plotpoint}}
\put(1067.0,323.0){\rule[-0.200pt]{4.818pt}{0.400pt}}
\put(1067.0,323.0){\rule[-0.200pt]{4.818pt}{0.400pt}}
\put(1093,313){\usebox{\plotpoint}}
\put(1083.0,313.0){\rule[-0.200pt]{4.818pt}{0.400pt}}
\put(1083.0,313.0){\rule[-0.200pt]{4.818pt}{0.400pt}}
\put(1104,306){\usebox{\plotpoint}}
\put(1094.0,306.0){\rule[-0.200pt]{4.818pt}{0.400pt}}
\put(1094.0,306.0){\rule[-0.200pt]{4.818pt}{0.400pt}}
\put(1215,235){\usebox{\plotpoint}}
\put(1205.0,235.0){\rule[-0.200pt]{4.818pt}{0.400pt}}
\put(1205.0,235.0){\rule[-0.200pt]{4.818pt}{0.400pt}}
\put(1325.0,160.0){\usebox{\plotpoint}}
\put(1315.0,160.0){\rule[-0.200pt]{4.818pt}{0.400pt}}
\put(1315.0,161.0){\rule[-0.200pt]{4.818pt}{0.400pt}}
\end{picture}